\documentclass[
 reprint, superscriptaddress,
 amsmath,amssymb,
 aps,
prl,
]{revtex4-1}

\usepackage{graphicx}% Include figure files
\usepackage{dcolumn}% Align table columns on decimal point
\usepackage{bm}% bold math
\usepackage{xcolor}
\usepackage{verbatim}
\usepackage{float}
\usepackage{color,soul}

\newcommand{\Tc}{$T_c$}

\newcommand{\ie}{\textit{i.e.}}

\begin{document}

%\preprint{APS/123-QED}

\title{Point Node Gap Structure of Spin-Triplet Superconductor UTe$_2$}

\author{Tristin Metz}
\author{Seokjin Bae}
\affiliation{Maryland Quantum Materials Center, Department of Physics, University of Maryland, College Park, Maryland 20742, USA}

\author{Sheng Ran}
\affiliation{Maryland Quantum Materials Center, Department of Physics, University of Maryland, College Park, Maryland 20742, USA}
\affiliation{NIST Center for Neutron Research, National Institute of Standards and Technology, Gaithersburg, MD 20899, USA}

\author{I-Lin Liu}
\affiliation{Maryland Quantum Materials Center, Department of Physics, University of Maryland, College Park, Maryland 20742, USA}
\affiliation{NIST Center for Neutron Research, National Institute of Standards and Technology, Gaithersburg, MD 20899, USA}
\affiliation{Department of Materials Science and Engineering,
University of Maryland, College Park, MD 20742 USA}

\author{Yun Suk Eo}
\author{Wesley T. Fuhrman}
\affiliation{Maryland Quantum Materials Center, Department of Physics, University of Maryland, College Park, Maryland 20742, USA}

\author{Daniel F. Agterberg}
\affiliation{Department of Physics, University of Wisconsin, Milwaukee, Wisconsin 53201, USA}

\author{Steven Anlage}
\affiliation{Maryland Quantum Materials Center, Department of Physics, University of Maryland, College Park, Maryland 20742, USA}

\author{Nicholas P. Butch}
\affiliation{Maryland Quantum Materials Center, Department of Physics, University of Maryland, College Park, Maryland 20742, USA}
\affiliation{NIST Center for Neutron Research, National Institute of Standards and Technology, Gaithersburg, MD 20899, USA}

\author{Johnpierre Paglione}
\email{paglione@umd.edu}
\affiliation{Maryland Quantum Materials Center, Department of Physics, University of Maryland, College Park, Maryland 20742, USA}
\affiliation{Canadian Institute for Advanced Research, Toronto, Ontario M5G 1Z8, Canada}

\date{\today}

\begin{abstract}
Low-temperature electrical and thermal transport, magnetic penetration depth, and heat capacity measurements were performed on single crystals of the actinide superconductor UTe$_2$ to determine the structure of the superconducting energy gap.
Heat transport measurements performed with currents directed along both crystallographic $a$- and $b$-axes reveal a vanishingly small residual fermionic component of the thermal conductivity.
The magnetic field dependence of the residual term follows a rapid, quasi-linear increase consistent with the presence of nodal quasiparticles, rising toward the $a$-axis upper critical field where the Wiedemann-Franz law is recovered. Together with a quadratic temperature dependence of the magnetic penetration depth up to $T/T_c=0.3$, these measurements provide evidence for an unconventional spin-triplet superconducting order parameter with point nodes. Millikelvin specific heat measurements performed on the same crystals used for thermal transport reveal an upturn below 300~mK that is well described by a divergent quantum-critical contribution to the density of states (DOS). Modeling this contribution with a $T^{-1/3}$ power law allows restoration of the full entropy balance in the superconducting state and a resultant cubic power law for the electronic DOS below $T_c$, consistent with the point-node gap structure determined by thermal conductivity and penetration depth measurements.

%\begin{description}
%\end{description}
\end{abstract}

\pacs{Valid PACS appear here}

\maketitle

%\tableofcontents

The uranium based superconductors URhGe, UCoGe, and UGe$_2$ have generated much interest due to the unusual coexistence of a ferromagnetic ground state with superconductivity present in all three systems \cite{Aoki19_U_based_SC_review}. Recently, the discovery of superconductivity in UTe$_2$ \cite{Ran19} has stimulated a flurry of experimental studies on this new actinide superconductor \cite{Ran19,Aoki19,Ran19_Extreme,Niu19,Miyake19,Sundar19}. With a ground state close to ferromagnetic order as evidenced by quantum critical scaling of the magnetization \cite{Ran19} and magnetic fluctuations down to milliKelvin temperatures \cite{Sundar19}, the UTe$_2$ system adds the long-sought paramagnetic end member to the family of uranium-based ferromagnetic superconductors. 

The superconducting state of UTe$_2$ is quite exotic. Strong evidence for spin-triplet pairing is given by a lack of change in the NMR knight shift upon cooling into the superconducting state, and an upper critical field $H_{c2}$ that exceeds the paramagnetic limit for all field directions (\ie, by more than an order of magnitude for $H\parallel b$) \cite{Ran19}. The NMR 1/T$_1$ relaxation time and the heat capacity were also observed to exhibit power law behavior consistent with point nodes in the gap structure \cite{Ran19}. 
Furthermore, a re-entrant superconducting phase was shown to develop in extremely high magnetic fields \cite{Ran19_Extreme}. 
Intriguingly, heat capacity measurements in the superconducting state have consistently exhibited an apparent residual fermionic ($T$-linear) term that appears to equal $\sim$50\% of the normal state electronic DOS \cite{Ran19,Aoki19}, which has been interpreted as a signature of an unpaired fluid.
However, this picture is problematic since a non-unitary pairing state is not formally possible given the orthorhombic crystal symmetry. More important, the entropy balance between normal and superconducting states that is required for a second-order transition (\ie, $\int_0^{T_c}\frac{C_{el_s}}{T}dT\ne\int_0^{T_c}\gamma_ndT$, where $C_{el_s}$ is the electronic heat capacity in the superconducting state and $\gamma_n$ is the normal state electronic heat capacity) is not obeyed \cite{Ran19,Aoki19}.

Here we use thermal conductivity and penetration depth measurements to probe the superconducting gap structure directly by observing the properties of low-energy quasiparticle excitations.
Thermal conductivity $\kappa$ has been established as a powerful directional tool for probing the gap structure of unconventional superconductors \cite{Shakeripour09}, and magnetic penetration depth is a well established tool for finding detailed information about the pairing state \cite{Prozorov2006SST}. In our thermal transport measurements, we observe a residual quasiparticle contribution which is vanishingly small in the zero field limit, but is restored by magnetic field much more quickly than would be expected for either clean or dirty limit fully gapped superconductivity. Together with a quadratic temperature dependence of the measured magnetic penetration depth, our observations are consistent with a point node superconducting gap structure in UTe$_2$.
We also present measurements of the specific heat of the same samples used for thermal transport in order to probe the reported residual DOS in the superconducting state. We observe a low-temperature upturn in the $T$-linear term  that is not consistent with a nuclear Schottky contribution, but rather a divergent term consistent with the presence of quantum critical degrees of freedom. Upon subtraction of the divergent term, we recover a complete entropy balance between superconducting and normal states in the electronic specific heat and the previously observed $T^3$ dependence \cite{Ran19,Aoki19}.

%%%%%%%%%%%%%%%%%%%%%%%%%%%%%%%%%%%%%%%%
\begin{table}[t]
    \centering
    \begin{tabular}{ccc}
    \hline
    \hline
         Representation & Representative Function & Nodes  \\
         \hline
         $A_{1g}$ & const$ + c_xkx^2 +c_y k_y^2+c_z k_z^2$ & none\\
         $B_{1g}$ & $c k_zk_y$ & lines\\ 
         $B_{2g}$ & $ck_zk_x$ & lines\\
         $B_{3g}$ & $ck_xk_y$ & lines\\
         $A_{1u}$ & $c_xk_x\mathbf{\hat{x}}+c_yk_y\mathbf{\hat{y}}+c_zk_z\mathbf{\hat{z}}$ & none \\
         $B_{1u}$ & $c_1k_y\mathbf{\hat{z}}+ c_2k_x\mathbf{\hat{y}}$ & points\\
         $B_{2u}$ & $c_1 k_x\mathbf{\hat{z}}+c_2k_z\mathbf{\hat{x}}$ & points\\
         $B_{3u}$ & $c_1k_x\mathbf{\hat{y}}+c_2k_y\mathbf{\hat{x}}$ & points\\
    \hline
    \hline
    \end{tabular}
    \caption{Candidate superconducting pairing states for the $D_{2h}$ point group symmetry.}
    \label{Gap Functions}
\end{table}

%METHODS...
Single crystals of UTe$_2$ were synthesized using a vapor transport method as described in \cite{Ran19}, using depleted uranium ($>$99.9\% $^{238}$U) and natural abundance elemental $^{127.60}$Te (99.9999\% purity). Crystals were shaped into rectangular plates with sample dimensions 1.5$\times$0.5$\times$0.1~mm$^3$ for the current $j\parallel a$ sample (S1) and  1.8$\times$1.0$\times$0.3~mm$^3$ for the $j\parallel b$ sample (S2), used for both thermal conductivity and specific heat measurements.
Thermal conductivity $\kappa$ was measured in a dilution refrigerator using a modular one-heater, two-thermometer probe \cite{Tanatar18}, with contacts made by soldering gold wire leads with an In-Sn alloy to evaporated gold pads, resulting in $\sim$m$\Omega$ contact resistances. 
Electrical resistivity $\rho$ was measured in-situ using the same wires and contacts, with samples in the exact position used for thermal transport measurements. Both samples exhibit a residual resistance ratio $\rho(300$~K)/$\rho(0$~K) of 22. As discussed below, the observation of the Wiedemann-Franz (WF) law in the field-induced normal state confirms the quality of measurements and rules out electron-phonon decoupling issues \cite{Smith05}.
Measurements of the low temperature magnetic penetration depth ($\lambda$) were performed using a dielectric resonator technique \cite{SeokjinBae2019RSI} on two other crystals, labeled S3 and S4, and heat capacity $C$ of samples S1 and S2 was measured using the relaxation time method. 

%THEORY...
The superconducting state in UTe$_2$ has been proposed \cite{Ran19,Ran19_Extreme,Aoki19} to be spin-triplet based on several experimental observations. Theoretically, because UTe$_2$ has an orthorhombic lattice that is symmorphic as confirmed at low temperatures by neutron scattering \cite{Hutanu19}, Blount's theorem must hold \cite{Blount85} so that odd-parity (\ie, spin-triplet) states will either have point nodes or be fully gapped. Conversely, even-parity (spin-singlet) states will have either line nodes or be fully gapped. More formally, assuming strong spin-orbit coupling and invoking the $D_{2H}$ point group yields the possible gap symmetries shown in Table~1. The possible symmetry-imposed nodal structures fall into three types: fully gapped ($A_{1g}$, $A_{1u}$), axial point nodes ($B_{1u}, B_{2u}, B_{3u}$ with pairs of point nodes along a single high-symmetry axis), or polar line nodes ($B_{1g}, B_{2g}, B_{3g}$ with lines node perpendicular to a single high symmetry axis). 

%%%%%%%%%%%%%%%%%%%%%%%%%%%%%%%%%%%%%%%%
\begin{figure}[t]
    \centering
    \includegraphics[width=0.5\textwidth]{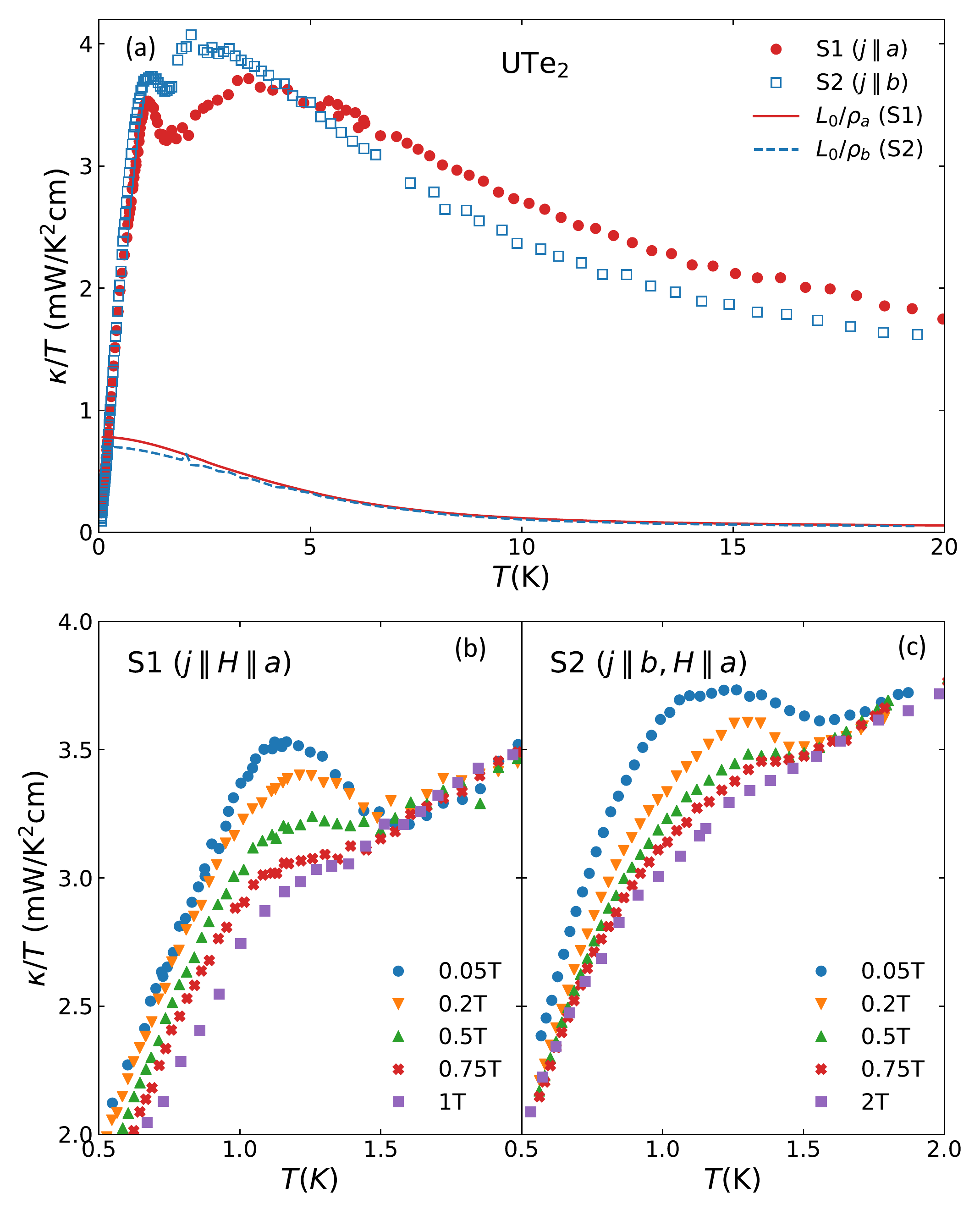}
    \caption{Thermal conductivity of UTe$_2$ samples S1 and S2.
    (a) Measured thermal conductivity in comparison to the estimated (elastic) charge contribution $L_0/\rho$ (solid and dashed lines) determined using the Wiedemann-Franz law and in-situ measured electrical resistivities of both samples. $L_0/\rho(T)$ is extrapolated below $T_c$ here assuming the resistivity follows the Fermi liquid temperature dependence $\rho = \rho_0+AT^2$. 
    Panels (b) and (c) present signatures of the superconducting transition in thermal conductivity measurements for (b) $j\parallel a$ and (c) $j\parallel b$ orientations. The observed increase in $\kappa/T$ below \Tc\ is thought to arise from a decrease in scattering due to the opening of the superconducting gap, and the suppression with increasing magnetic fields due to vortex scattering.}
    \label{fig:kappa_jump_near_Tc}
\end{figure}

%KAPPA
Thermal conductivity is a directional probe that can be used to determine the momentum space position of nodes by studying low-energy quasiparticle excitations -- as a function of field angle, as shown in the cuprate superconductor YBCO  \cite{Yu95,Aubin97} and heavy fermion compound CeCoIn$_5$ \cite{Izawa01}, or as a function of the directional current as shown in UPt$_3$ \cite{Suderow97,Lussier96} and CeIrIn$_5$ \cite{Shakeripour07}.
Furthermore, the magnetic field evolution of the electronic thermal conductivity in the $T=0$ limit ($\kappa_0(H)$) is a reliable probe of both nodal gaps (e.g., UPt$_3$\cite{Suderow97} and KFe$_2$As$_2$ \cite{Reid12}) and multi-band or anisotropic gap superconductors \cite{Boaknin03}.

%%%%%%%%%%%%%%%%%%%%%%%%%%%%%%%%%%%%%%%%
\begin{figure*}[t]
    \centering
    \includegraphics[width=\textwidth]{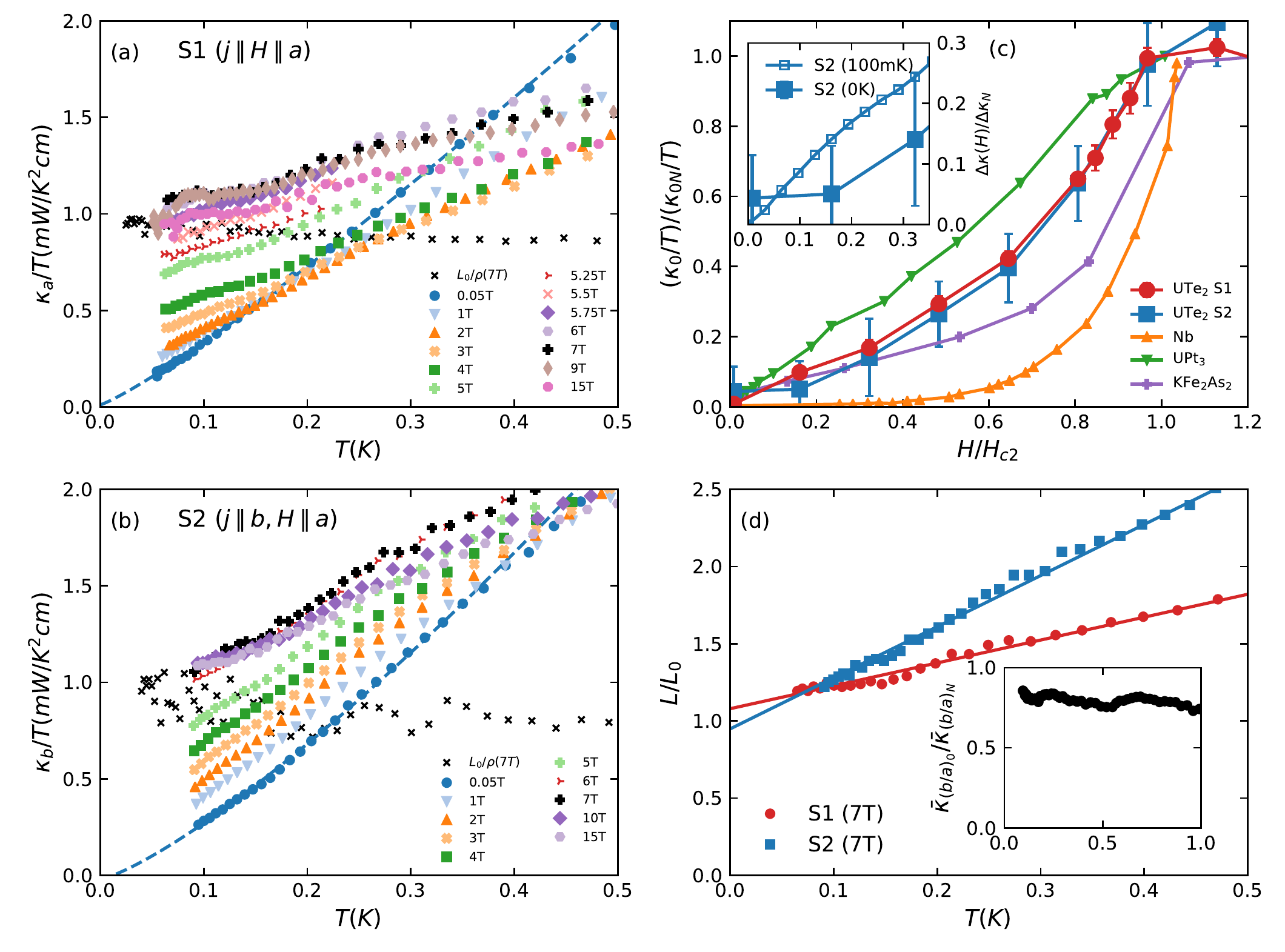}
    \caption{Low temperature thermal conductivity of UTe$_2$ single crystal samples S1 (a) and S2 (b), measured with heat current applied along crystallographic $a-$ and $b$-axes, respectively, and magnetic fields applied along the $a$-axis. 
    The black $\times$ in (a) and (b) represent the converted normal state charge conductivity $L_0/\rho$ measured in-situ at 7~T and calculated using the Wiedemann-Franz law. (c) Magnetic field dependence of $T\to 0$~K extrapolated values (, $\kappa_0/T \equiv \kappa(T\to 0)/T$) taken from the data in (a) and (b). Error bars correspond to uncertainty in the extrapolation from fitting to a power law over different temperature windows. For comparison we include data for elemental clean-limit fully gapped superconductor Nb \cite{Lowell70} and clean-limit nodal superconductors UPt$_3$ \cite{Suderow97} and KFe$_2$As$_2$ \cite{Reid12}. The inset shows a comparison of the detailed low field dependence of $\Delta\kappa(H)/\Delta\kappa_N\equiv(\kappa(H)-\kappa(0))/(\kappa(7$~T)$-\kappa(0))$ measured in field-sweep mode at 100mK, with the $T\to 0$ extrapolated values from the main panel. (d) Lorenz ratio $L/L_0$ calculated using the ratio of measured quantities $\kappa \rho /T$ at 7~T from (a) and (b) to the Lorenz number $L_0$. Linear extrapolations to $T=0$ show agreement with $L_0$ to within 8\% and 5\% for $j\parallel a$ and $j\parallel b$, respectively. The inset shows the zero-field anisotropy ratio $\bar{\kappa}=\kappa_b/\kappa_a$ normalized to the anisotropy ratio in the normal state $\bar{\kappa}_N=\bar{\kappa}(7$~T).}
    \label{fig:kappaonT_lowT_in_field}
\end{figure*}

Fig.~\ref{fig:kappa_jump_near_Tc}(a) presents the total thermal conductivity of UTe$_2$ samples S1 and S2 measured up to 20~K and compared to the estimated (elastic) charge contribution determined using the in-situ measured resistivity $\rho$ for each sample and the WF law (\ie, $L_0/\rho$, where $L_0=\pi^2/3(k_B/e)^2$),  showing that the majority of the thermal conduction in this range is not electronic. This makes distinguishing the quasiparticle contribution to thermal conductivity quite difficult, even within the superconducting regime.
For both samples, there is an abrupt increase in $\kappa/T$ at $T_c$ as highlighted in Figs.~\ref{fig:kappa_jump_near_Tc} (b) and (c), similar to that observed in CeCoIn$_5$ \cite{Mosovich01,Tanatar05} and YBCO \cite{Cohn92,Yu92}, where it is attributed to a suppression of the scattering of heat carriers (electrons and phonons, respectively) by the opening of the superconducting gap. In the case of UTe$_2$, the enhanced conductivity below $T_c$ is likely due to a combination of electron and phonon carriers being affected by the gap opening, and, more important, is an indication of the high crystalline quality of the samples being measured. The rapid suppression of the enhancement in field is likely due to vortex scattering.

The nature of the superconducting gap structure can be determined by probing the value of the residual term of $\kappa/T$ in the $T=0$ limit, $\kappa_0/T$, and its magnetic field dependence \cite{Graf96}. First, we confirm that the WF law is obeyed in the $T$=0 limit in the field-induced normal state at 7 T ($a$-axis). To study the zero-field limit, we apply a small field of 50~mT to ensure soldered contacts are not superconducting and extrapolate to $T=0$ by fitting the lowest temperature data to $\kappa/T = \kappa_0/T+AT^\alpha$. 
As shown in Fig.~\ref{fig:kappaonT_lowT_in_field} (a) and (b), we find that $\kappa/T$ extrapolates to a vanishingly small residual value for both heat current orientations and rapidly rises with temperature. This is inconsistent with the behavior of a fully gapped superconductor \cite{BRT59}, which would not conduct heat until at least $\sim 20\%$ of $T_c$, and is more consistent with a gap with nodal excitations \cite{Graf96}.

To rule out the possibility of a strongly anisotropic but fully gapped scenario, we study the field dependence of $\kappa_0/T$. 
For a full gap,  quasiparticles remain localized to vortex cores and only contribute to the thermal conductivity for fields exponentially close to $H_{c2}$, such as shown for the classic experiment on elemental Nb shown in Fig.~\ref{fig:kappaonT_lowT_in_field}(c) \cite{Lowell70}. For a fully gapped superconductor in the dirty limit, $\kappa_0/T$ increases more quickly in field but still remains exponentially small in the low-field limit \cite{Willis76}).
In multi-band superconductors with a disparity in gap values, a rapid rise can occur due to a very small field range $H^*$ of activated behavior of a small gap $\Delta_1$ (\ie, where $\Delta_1/\Delta_2 \sim \sqrt{H^*/H_{c2}}$ 
\cite{Boaknin03,Yamashita09}). For a nodal gap structure, low energy quasiparticle excitations are not bound and therefore follow a much stronger field dependence that depends on the specific gap structure (e.g. for line nodes, $\kappa_0/T \propto\sqrt{H}$ due to the Volovik effect).

Using the same extrapolation method as above, we plot in Fig.~\ref{fig:kappaonT_lowT_in_field}(c) the field dependence of 
$\kappa_{0a}/T$ and $\kappa_{0b}/T$, which both exhibit a rapid quasi-linear rise with magnetic field, further confirmed by detailed 100~mK field-sweep measurements shown in the inset of Fig.~\ref{fig:kappaonT_lowT_in_field}(c). 
This behavior rules out a fully gapped scenario, including clean limit and dirty limit behaviors, and places a strong constraint on the presence of a very small gap (e.g. $\Delta_1 \ll \Delta_2$).
Furthermore, we note that the temperature dependence of $\kappa/T$ in the presence of such a small gap should show a very rapid increase at low temperatures (or even an apparent residual term), making the extreme multi-band scenario unlikely.
Rather, the field dependence of $\kappa_0/T$ in UTe$_2$ is more reminiscent of that of nodal gap superconductors. Similar behavior was observed in line-node superconductors CeIrIn$_5$ \cite{Shakeripour09}, Sr$_2$RuO$_4$ \cite{Tanatar01}, and KFe$_2$As$_2$ \cite{Reid12}, which are all accompanied by finite residual values in zero field. 
A line-node (e.g. $d$-wave) gap is expected to invoke a universal zero-field residual term given by $\kappa_{0}/T=\frac{1}{3}\gamma_N v_F^2\frac{a\hbar}{2\mu \Delta_0}$. Using a conservative upper limit of $\kappa_0/T$=0.1~mW/cm/K$^2$ from our experiment together with values $2\Delta_0=3.5 k_B T_c$, $\gamma_N$=100 mJ/mol/K$^2$ and $\mu=2$ for a standard $d$-wave gap slope at the nodes ($a$ is order unity) yields $v_F$=6,000~m/s, a very small velocity similar to that of the extremely heavy $p$-wave superconductor UPt$_3$ \cite{TailleferRMP02}, and likely not consistent with UTe$_2$.
Interestingly, UPt$_3$ exhibits both vanishingly small $\kappa_0$ in zero field \cite{TailleferRMP02} and a linear increase with magnetic field \cite{Suderow97}, also shown for comparison in Fig.~\ref{fig:kappaonT_lowT_in_field}(c). 
Together with this striking similarity, the vanishingly small residual $T$-linear term and rapid quasi-linear rise in magnetic field observed for  UTe$_2$ are most consistent with a point-node gap scenario. 

Further information can be garnered from the anisotropy ratio $\bar{\kappa}\equiv\kappa_{b}/\kappa_{a}$, shown in the inset of Fig.~\ref{fig:kappaonT_lowT_in_field}(d). Normalizing out the normal state anisotropy by using the ratio $\bar{\kappa}(0.05$~T$)/\bar{\kappa}(7$~T), we find the superconducting state anisotropy approaches $\sim$0.8 in the low-$T$ limit. This not as large as the value ($\sim 0.5$) observed in UPt$_3$ \cite{Lussier96}, a more compelling indicator of point node direction. Rather, the near-unity anisotropy may arise due to nodal excitations directed in both basal plane directions (e.g. accidental), but will require detailed calculations based on candidate gap structures to draw firm conclusions.

\begin{figure}
    \centering
    \includegraphics[width=\columnwidth]{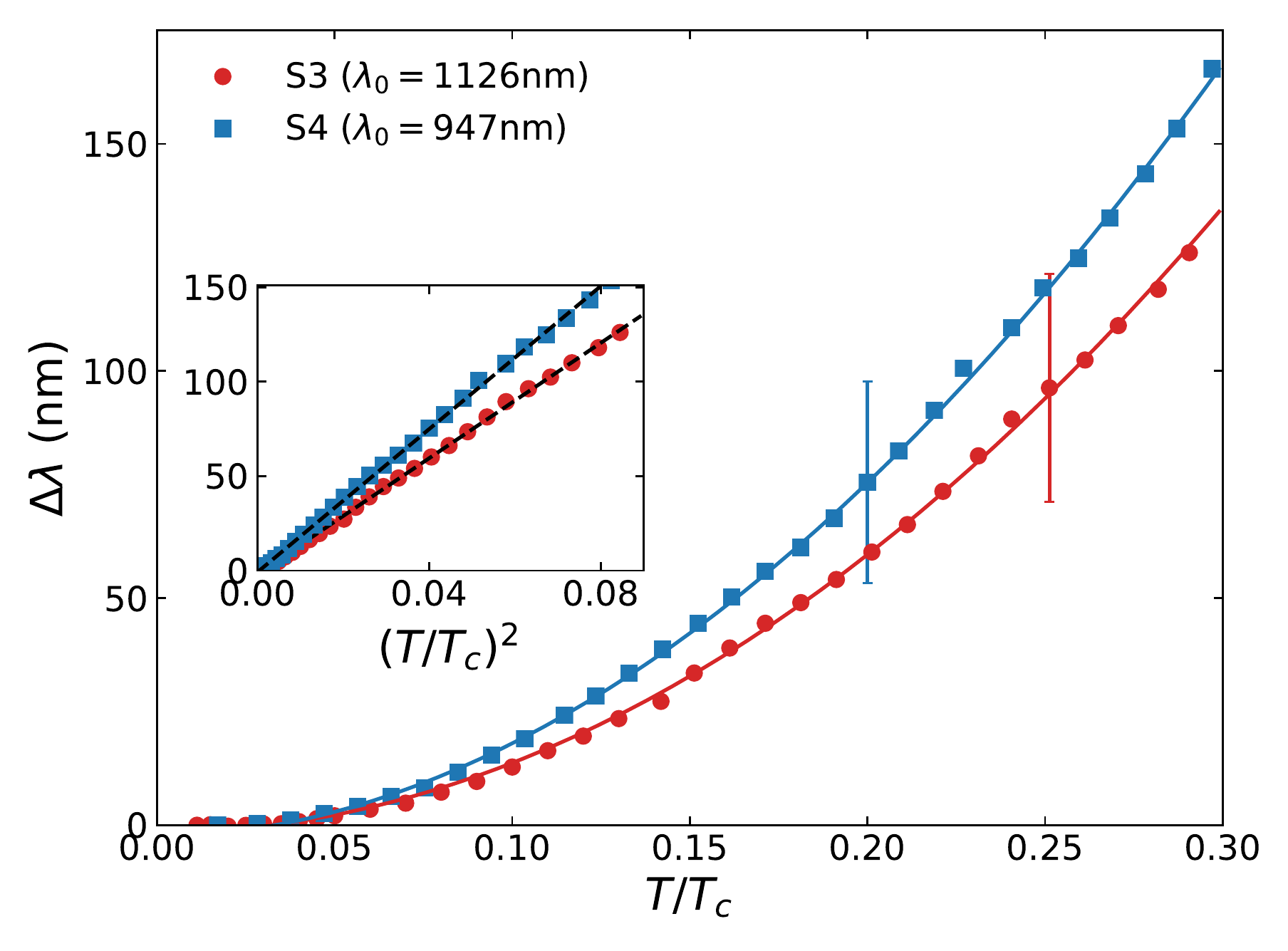}
    \caption{Effective penetration depth of UTe$_2$ samples S3 (red circles) and S4 (blue squares) plotted vs normalized temperature. The solid lines are power law fits of the form $aT^n+b$, with extracted exponents $n=2.00$ for S3 and $n=1.92$ for S4. The vertical bars shown for each data set are the conservative uncertainty estimates. The inset presents the same data in terms of $(T/T_c)^2$ to illustrate the quadratic temperature dependence, with linear dasned lines as guides.}
    \label{fig:lowTlambda_UTe2}
\end{figure}

%% PENETRATION DEPTH
Measurements of the low temperature magnetic penetration depth were performed using a cylindrical rutile dielectric resonator \cite{Bae2019arXiv2}, which facilitates a microwave transmission resonance involving the $ab$-plane electrodynamic response at $\approx 11$~GHz. The temperature dependence of the resonant frequency $\Delta f_0(T)$ and the quality factor $Q(T)$ are measured and converted into surface impedance $Z_s$ through a geometrical factor estimated from the geometry of the resonator and the nature of the mode \cite{Hein1999}.  In the local electrodynamic limit, where the mean free path is smaller than the screening length, $Z_s$ can be converted to the complex conductivity $\tilde{\sigma}=\sigma_1 - i\sigma_2$ of the samples through the local relation $Z_s = \sqrt{i\mu_0\omega/\tilde{\sigma}}$ \cite{Dressel2002}. The effective penetration depth is obtained from the imaginary part of the conductivity $\sigma_2 = 1/\mu_0\omega\lambda^2_{eff}(T)$, which yields $\lambda_{eff}(0)=1126$ nm for S3 and 947 nm for S4, both similar to the estimation ($\lambda(0)\gtrsim 1000$ nm) from a recent study \cite{Sundar19}. 

As shown in Fig.~\ref{fig:lowTlambda_UTe2}, the change in screening length, $\Delta\lambda_{eff}(T)$, exhibits a rapid rise with temperature indicative of low-lying excitations  \cite{Prozorov2006SST}. 
Fitting $\Delta\lambda_{eff}(T)$ to a power law form $aT^n+b$ below $0.3 T_c$ yields a quadratic temperature exponent ($n=2.00 \pm 0.01$ for S3 and $n=1.92 \pm 0.01$ for S4) consistent with either a dirty limit line node \cite{Hirschfeld1993PRB} or axial point node scenario \cite{Einzel1986PRL, Gross1986ZPhyB}. Considering the observed vanishingly small $\kappa_0/T$ values, and the deduced clean limit of the superconducting state ( mean-free-path of 95~nm estimated from the measured scattering rate $\Gamma\leq0.005 k_BT_c$ \cite{Bae2019arXiv2}) being much larger than the superconducting coherence length ($4\sim 7$ nm) deduced from $H_{c2}$ measurement \cite{Ran19}), we exclude the dirty limit scenario for this system and conclude that the quadratic dependence of $\Delta\lambda_{eff}(T)$ is more consistent with axial point nodes aligned in the direction of the vector potential. In our setup, the applied microwave magnetic field generates circulating screening currents in the $ab$-plane \cite{SeokjinBae2019RSI}, so that the direction of axial point nodes is assumed to lie in the $ab$-plane.

\begin{figure}[h!]
    \centering
    \includegraphics[width=0.49\textwidth]{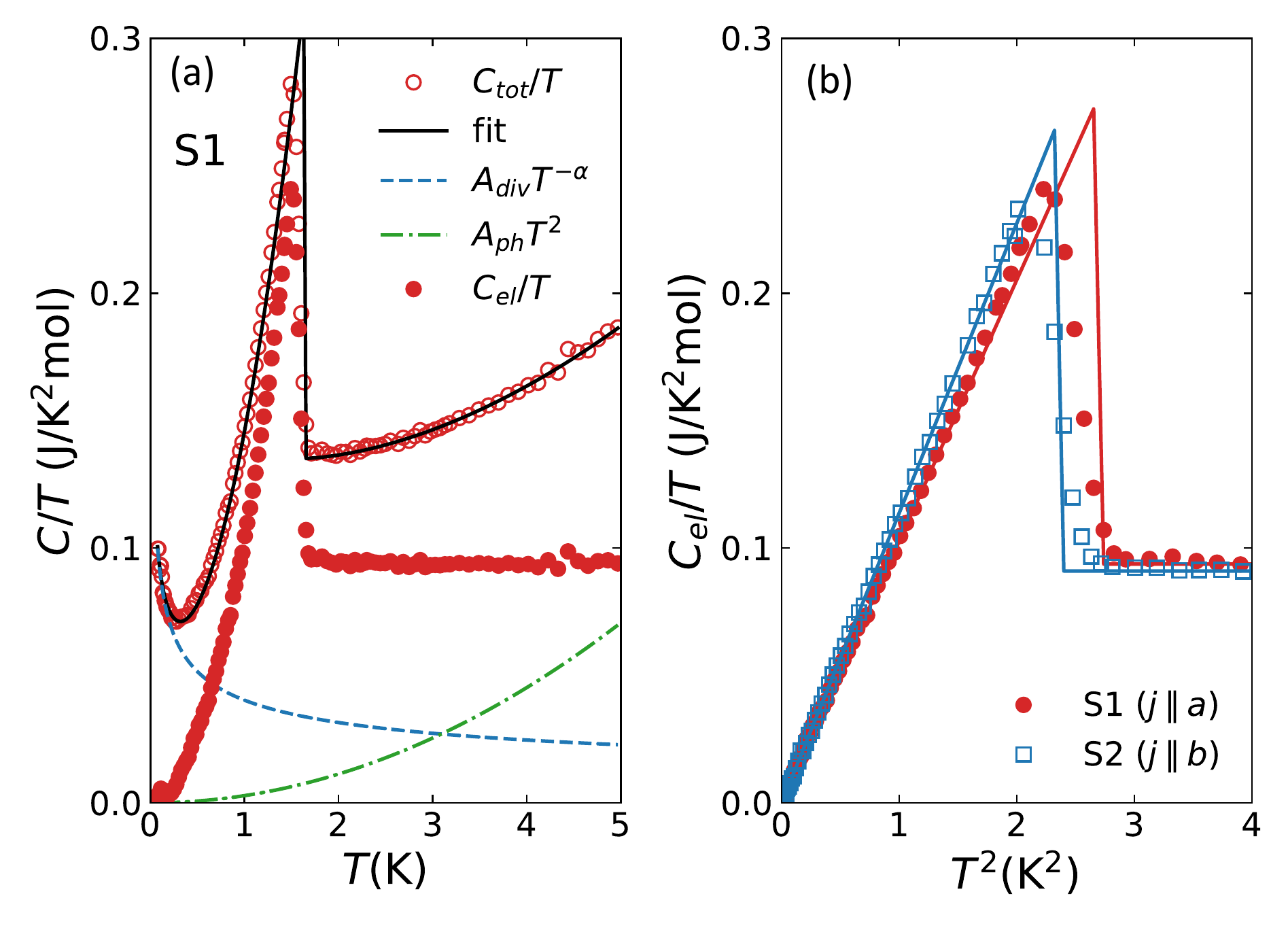}
    \caption{Analysis of the low temperature heat capacity of UTe$_2$ single-crystal samples S1 and S2 (same crystals used for thermal transport measurements).
    (a) The total measured heat capacity (open circles) of sample S1 decomposed into a weak power law divergence (dashed line), phonon term (dash-dotted line), and electronic heat capacity (filled circles) obtained by subtracting the diverging and phonon terms (see text for details). (b) The obtained electronic heat capacities of samples S1 and S2 plotted vs. $T^2$. Solid lines are fits to the $T^2$ dependence in the superconducting state with $T_c$ determined by balancing entropy.}
    \label{fig:50mT_data}
\end{figure}

%SPECIFIC HEAT
Concerning previous observations of an apparent residual DOS in specific heat \cite{Ran19,Aoki19}, our observations of a vanishingly small fermionic component of thermal conductivity in the zero-field, zero-temperature limit are inconsistent with a scenario that entails the presence of a sizeable residual non-paired fluid. To confirm this, Fig.~\ref{fig:50mT_data} presents the measured specific heat of samples S1 and S2, which are the same crystals used for thermal transport. 
This measurement reproduces the large residual value in $C/T$ observed in previous experiments \cite{Ran19,Aoki19}, but uncovers a sharp increase in $C/T$ below about 300~mK. Although it appears similar to a nuclear Schottky anomaly, which arises from the splitting of nuclear spin states, such a contribution ($C_N$) should be negligible since the only isotope with a nuclear spin is $^{125}$Te, which is only 7\% abundant in natural Te and does not have a nuclear quadrupole moment. Moreover, the observed upturn is much broader than the expected form $C_N/T\propto T^{-3}$. Instead, it is well fit by a weak power law ($C/T\propto T^{-\sim1/3}$) divergence as described below. 

Motivated by this, we fit the total heat capacity over the entire temperature range (\ie, 50~mK to 5~K) using
\begin{equation}
    \frac{C}{T}=A_{ph}T^2+A_{div}T^{-\alpha}+\begin{cases}A_{sc}T^2&T<T_c\\\gamma_n&T>T_c\end{cases}.
\end{equation}
The result is shown in Fig.~\ref{fig:50mT_data}(a), where the solid line represents the total fit, $A_{ph}T^2$ is the phonon heat capacity (dash-dotted line), $A_{div}T^{-\alpha}$ is the power law divergence (dashed line) where $\alpha$ = 0.35 (0.33) for S1 (S2), and the last term is the electronic heat capacity which varies as $T^2$ in the superconducting regime ($T<T_c$) and is constant ($\gamma_n$) above $T_c$. 

The extracted electronic component $C_{el}/T=C/T-A_{ph}T^2-A_{div}T^{-\alpha}$ exhibits a nearly perfect entropy balance between normal and superconducting states, which is reflected by the fit result in Fig.~\ref{fig:50mT_data}(a) where the resultant $T_c$ (set by the entropy balance) matches the data very well. Fig.~\ref{fig:50mT_data}(b) shows that the specific heat for both S1 and S2 samples follows a cubic power law (\ie, $C_{el}/T \sim T^2$) through the superconducting temperature range, consistent with prior fits to raw data above 300~mK \cite{Ran19,Aoki19}. 
While the divergent residual specific heat term requires further investigation, the lack of any sizeable contribution to zero-field thermal conduction means that this component arises from either a well-localized DOS, or itinerant fermionic carriers that are heavily scattered in the $T \to 0$ limit.

In conclusion, detailed measurements of thermal conductivity, penetration depth and heat capacity in UTe$_2$ provide evidence for a superconducting gap structure with point nodes. The vanishingly small residual fermionic term in thermal conductivity, together with rapid increases with both temperature and magnetic field in the zero-field, zero-temperature limits, respectively,  confirm the presence of low-energy quasiparticle excitations and the proposed nodal gap structure, and rule out an anistropic or multi-band full gap scenario. A quadratic temperature dependence of the low temperature penetration depth is also consistent with an nodal gap structure, with point nodes in the $ab$-plane.
Together with consideration of the $D_{2H}$ point group symmetry, these results provide evidence for spin-triplet Cooper pairing in UTe$_2$.
The origin of a localized or strongly scattered divergent quantum critical component of the specific heat, which underlies the superconducting state, will require further study to elucidate.


\begin{thebibliography}{99}
\bibitem{Aoki19_U_based_SC_review}
D. Aoki \textit{et al.}, J. Phys. Soc. Jpn. \textbf{88}, 022001 (2019).

\bibitem{Ran19}
S. Ran \textit{et al.}, Science \textbf{365}, 684 (2019).

\bibitem{Aoki19}
D. Aoki \textit{et al.}, J. Phys. Soc. Jpn. \textbf{88}, 043702 (2019).

\bibitem{Ran19_Extreme}
S. Ran \textit{et al.}, Nat. Phys. (2019), doi:10.1038/s41567-019-0670-x.

\bibitem{Niu19}
Q. Niu \textit{et al.}, arxiv:1907.11118.

\bibitem{Miyake19}
A. Miyake \textit{et al.}, J. Phys. Soc. Jpn. \textbf{88}, 063706 (2019).

\bibitem{Sundar19}
S. Sundar \textit{et al.}, Phys. Rev. B \textbf{100}, 140502(R) (2019).

\bibitem{Shakeripour09}
H. Shakeripour \textit{et al.}, New J. Phys. \textbf{11}, 055065 (2009).

\bibitem{Prozorov2006SST}
R. Prozorov and R.W. Giannetta, Supercond. Sci. Technol. \textbf{19}, R41 (2006).

\bibitem{Tanatar18}
M.A. Tanatar \textit{et al.}, Rev. Sci. Instr. \textbf{89}, 013903 (2018).

\bibitem{Smith05}
M. Smith \textit{et al.}, Phys. Rev. B \textbf{71}, 014506 (2005).

\bibitem{SeokjinBae2019RSI}
S. Bae \textit{et al.}, Rev. Sci. Instrum. \textbf{90}, 043901 (2019).

\bibitem{Hutanu19}
V. Hutanu \textit{et al.}, arxiv:1905.04377.

\bibitem{Blount85}
E.I. Blount, Phys. Rev. B \textbf{32}, 2935 (1985).

\bibitem{Yu95}
F. Yu \textit{et al.}, Phys. Rev. Lett. \textbf{74}, 5136 (1995).

\bibitem{Aubin97}
H. Aubin \textit{et al.}, Phys. Rev. Lett. \textbf{78}, 2624 (1997).

\bibitem{Izawa01}
K. Izawa \textit{et al.}, Phys. Rev. Lett. \textbf{87}, 057002 (2001).

\bibitem{Suderow97}
H. Suderow \textit{et al.}, J. Low Temp. Phys. \textbf{108}, 11 (1997).

\bibitem{Lussier96}
B. Lussier \textit{et al.}, Phys. Rev. B \textbf{53}, 5145 (1996).

\bibitem{Shakeripour07}
H. Shakeripour \textit{et al.}, Phys. Rev. Lett. \textbf{99}, 187004 (2007).

\bibitem{Reid12}
J.-Ph. Reid \textit{et al.}, Phys. Rev. Lett. \textbf{109}, 087001 (2012).

\bibitem{Boaknin03}
E. Boaknin \textit{et al.}, Phys. Rev. Lett. \textbf{90}, 117003 (2003).

\bibitem{Mosovich01}
R. Mosovich \textit{et al.}, Phys. Rev. Lett. \textbf{86}, 5152 (2001).

\bibitem{Tanatar05}
M. Tanatar \textit{et al.}, Phys. Rev. Lett. \textbf{95}, 067002 (2005).

\bibitem{Cohn92}
J.L. Cohn \textit{et al.}, Phys. Rev. B \textbf{45}, 1314(R) (1992).

\bibitem{Yu92}
R.C. Yu \textit{et al.}, Phys. Rev. Lett. \textbf{69}, 1431 (1992).

\bibitem{Graf96}
M. J. Graf \textit{et al}., Phys. Rev. B \textbf{53}, 15147 (1996).

\bibitem{BRT59}
J. Bardeen \textit{et al.}, Phys. Rev. \textbf{113}, 982 (1959).

\bibitem{Lowell70}
J. Lowell \textit{et al.}, J. Low Temp. Phys. \textbf{3}, 65 (1970).

\bibitem{Willis76} %%Indium-Bismuth study
J.O. Willis and D.M. Ginsberg, Phys. Rev. B \textbf{14}, 1916 (1976).

\bibitem{Yamashita09}
M. Yamashita \textit{et al.}, Phys. Rev. B \textbf{80}, 220509(R) (2009).

\bibitem{Tanatar01}
M. Tanatar \textit{et al.}, Phys. Rev. B \textbf{63}, 064505 (2001).

\bibitem{TailleferRMP02}
R. Joynt and L. Taillefer, Rev. Mod. Phys. \textbf{74}, 235 (2002).

\bibitem{Bae2019arXiv2}
S. Bae \textit{et al.}, arXiv:1909.09032.

\bibitem{Hein1999}
M. A. Hein, \textit{High-Temperature Superconductor Thin Films at Microwave Frequencies} (Springer, Heidelberg, 1999) pp. 45-46.

\bibitem{Dressel2002}
M. Dressel and G. Gruner, \textit{Electrodynamics of Solids} (Cambridge University Press, 2002).

\bibitem{Hirschfeld1993PRB}
P. J. Hirschfeld and N. Goldenfeld, Phys. Rev. B \textbf{48}, 4219 (1993).

\bibitem{Einzel1986PRL}
D. Einzel \textit{et al.}, Phys. Rev. Lett. \textbf{56}, 2413 (1986).

\bibitem{Gross1986ZPhyB}
F. Gross \textit{et al.}, Z. Phy. B Cond. Matt. \textbf{64}, 175 (1986).
\end{thebibliography}
\end{document}